\documentclass[preprint,amsmath,amssymb]{revtex4}
\input epsf
\usepackage{color}

\def\be{\begin{equation}}
\def\ee{\end{equation}  }
\def\bea{\begin{eqnarray}}
\def\eea{\end{eqnarray}  }

\begin{document}
\title{Dark Energy, Inflation and Extra Dimensions}

\author{Paul J. Steinhardt$^{1,2}$ and Daniel Wesley$^3$}
\affiliation{ $^1$Joseph Henry Laboratories, 
Princeton University, Princeton, NJ 08544 \\
$^2$Princeton Center for Theoretical Science, 
Princeton University, Princeton, NJ 08544 \\
$^3$Centre for Theoretical Cosmology, DAMTP, Cambridge University, 
Wilberforce 
Road, Cambridge CB3 0WA, UK }

\begin{abstract}
We consider how accelerated expansion, whether due to inflation or 
dark energy, imposes strong constraints on fundamental theories 
obtained by compactification from higher dimensions.  For 
theories that obey the null energy condition (NEC), we find that 
inflationary cosmology is impossible 
for a wide range of compactifications; and 
a dark energy phase consistent with observations is only possible if 
both Newton's gravitational constant 
and the dark energy equation-of-state 
vary with time.  
If the theory violates the NEC, inflation and dark energy are only 
possible if the 
NEC-violating elements are inhomogeneously distributed in the 
compact dimensions and vary with time in precise synchrony with the 
matter and energy density in the non-compact dimensions. 
Although our proofs are derived assuming 
 general  relativity applies in both four and higher dimensions and certain forms of metrics, we argue that 
 similar constraints must 
apply for more general compactifications.
\end{abstract}

\maketitle

\section{Introduction}

Compelling evidence exists that the present universe is dominated by 
some form of dark energy and undergoing a period of accelerated 
expansion.  Also, a widely accepted hypothesis is that the early 
universe underwent inflation, a period of accelerated expansion 
shortly after the big bang that smoothed and flattened the universe 
and generated a nearly scale-invariant spectrum of density 
perturbations.  

The purpose of this paper is to explore the implications of cosmic 
acceleration for fundamental theories obtained by compactification 
from a higher dimensional theory, a feature common to Kaluza-Klein 
theory, Randall-Sundrum models, string theory and M-theory, for 
example.   
A general property of compactified theories is that the expansion of 
the non-compact directions required for 
any realistic big bang cosmology has the tendency to cause the extra 
dimensions to contract unless some interaction prevents it.  The 
contraction of the extra dimensions has undesirable physical
effects, such as the time variation of Newton's constant or other 
fundamental 
constants and a deviation from standard Friedmann-Robertson-Walker 
(FRW) evolution.  For decelerating universes, these problems can be 
avoided, in principle, by introducing ordinary interactions.  

In this paper, though, combining techniques developed in 
Refs.~\cite{Wesley:2008fg, Wesley:2008de} with new approaches, we 
shall derive a series of no-go theorems showing how one is forced to 
consider more exotic solutions  in order to obtain
accelerated expansion in
 compactified theories.  
The power of these theorems may surprise some readers, yet they emerge 
from fairly simple considerations.  The key constraint is that the
models are described by Einstein gravity both in the 4d  effective 
theory and in the higher dimensional theory.  What seems relatively 
innocuous in 
the  4d effective theory -- {\it e.g.,} accelerated expansion of the 
non-compact directions -- can require something extraordinary when 
lifted 
into the higher dimensional Einstein gravity.  As a simple example, 
consider the original
Kaluza-Klein model with a single static extra dimension 
whose size, we will assume, has been frozen by 
some interaction. 
Accelerated expansion of the 4d effective theory means that the 5d 
theory is described by a metric $ds^2 =- dt^2 +a^2 (dx_1^2+ dx_2^2 + 
dx_3^2) +dx_4^2$, where the FRW scale factor satisfies
$\dot{a}>0$ and $\ddot{a}>0$.
   By substituting the metric into the 5d 
Einstein equations, it is possible to show \cite{Wesley:2008fg} that 
the equation-of-state in 
the compact dimension 
(the ratio of the 4-4 to the 0-0 components
of the energy-momentum 
tensor) is $w_5<-1$;  for example, for an expanding universe with 
$a(t) \sim t^{p>1}$, its value is 
$w_5= (1-2 p)/p<-1$.  The fact that this 
ratio is less than -1 means the higher 
dimensional theory necessarily violates the null 
energy condition (NEC),
an extraordinary constraint. The NEC is not violated by any 
observed matter fields or by unitary two-derivative quantum field 
theories; 
and violating the NEC can produce problems of its own.  Under many 
conditions it leads to unacceptable consequences, such as superluminal 
propagation, instabilities, or violations of 
unitarity.\cite{NEC1,NEC2,NEC3,NEC4,Marvel:2008uh} 

We begin in Sec.~\ref{SecIII} by considering compactified theories 
that do not violate the NEC,  including
 the original
 Kaluza-Klein model, the Randall-Sundrum II model\cite{RS2}
 and many string theory models, and see 
how difficult it is to accommodate accelerated expansion. 
For a wide class of models, we derive 
a no-go theorem that rules out inflationary 
cosmology altogether and additional no-go theorems that rule out the 
simplest dark energy models, including $\Lambda$CDM.  We further show 
that a dark energy 
phase with accelerated expansion consistent with current observations 
is only possible if both Newton's gravitational constant and the dark 
energy equation-of-state vary with time.

Then, we turn our attention in Sec.~\ref{SecIV} to models that do 
violate the NEC. In spite of the potential dangers cited above and in 
Refs.~\cite{NEC1,NEC2,NEC3,NEC4,Marvel:2008uh}, models of this type 
have been 
suggested that may safely violate NEC, such as 
the Randall-Sundrum I model\cite{RS1}
and recently proposed flux 
compactifications on the string landscape.  
Examples of NEC-violating components invoked in string
constructions include orientifold-planes, which have negative 
tension, and quantum effects
analogous to Casimir energy.
Here, though, we find 
another set of new no-go theorems that rule out 
some forms of NEC violation and impose precise
conditions on how any NEC-violating components
must vary with time as the universe evolves.   
Although the discussion here is confined to certain
 common types of metrics and assumes Einstein's general theory of relativity applies in higher dimensions, 
 we argue in Sec.~V 
 that similar no-go theorems must apply
 in more general cases.

Our approach complements but is quite different 
from previous
no-go theorems based on supersymmetry or supergravity 
\cite{de Wit:1986xg}; 
supersymmetry is not assumed in our analysis, so our conclusions apply 
to more general compactified
theories.  Our results are also different from 
inflationary no-go theorems based on requiring small values of the 
slow-roll parameters $\epsilon$ and $\eta$ in the case of inflation; 
or  constructions 
leading to the long-lived metastable de Sitter minima in the string 
landscape 
\cite{Giddings:2001yu,DeWolfe:2002nn,Kachru:2003aw,Giddings:2005ff,Douglas:2006es}.  
Previous theorems are based on what 
might be called 
 ``micro-to-macro"
 approaches where 
the microphysics is specified first and then the constraints on the 
macroscopic pressure, 
energy density and equation-of-state are derived.

Ours is a more ``macro-to-micro" approach in which we assume a certain 
equation of state on macroscopic scales (based on observations) and 
derive constraints on 
the microphysics.  This method is more closely related to the one 
used 
by 
various authors \cite{Gibbons:1985,Maldacena:2000mw, 
Carroll:2001ih, Giddings:2003zw} to constrain compactified 
theories 
with purely static de Sitter minima (equation-of-state 
$w=w_{\rm{DE}}=-1$, 
where 
we use $w$ to represent the ratio of 
{\it total} pressure to {\it total} energy density 
and $w_{\rm{DE}}$ to represent the pressure-to-density
ratio for the dark energy component alone). 
In Refs. \cite{Wesley:2008fg,Wesley:2008de} and
this paper, though, the
 constraints are derived for more general -- and more 
practical -- cases where 
$w$ is significantly greater than $ -1$ and time-varying
({\it e.g.,} the present 
universe has $w \approx -0.74$ today and varying with time)
\cite{Riess:2006fw,Kowalski:2008ez,Komatsu:2008hk}.
 By considering the time-evolution 
in $w$,
we derive numerous new constraints
that do apply in the pure de 
Sitter limit, $w=-1$.  Another new feature of this paper is that it 
derives no-go theorems for a wide class of 
time-dependent metrics that were not constrained
previously (the ``CRF metrics" described below).

\section{Compactified Models and NEC violation} \label{SecII}

The NEC is commonly assumed in 
fundamental theories to avoid the 
classical and quantum instabilities (closed time-like curves, big 
rips, ghosts and  unitarity 
violation) normally associated with its violation
\cite{NEC1,NEC2,NEC3,NEC4,Marvel:2008uh}.   Nevertheless, we 
will show that, for a wide range of compactified models, inflationary 
cosmology and the NEC are completely incompatible and that dark energy 
is compatible only if Newton's gravitational constant $G_N$ and the 
dark 
energy equation-of-state $w_{\rm{DE}}$ vary with time. 

\subsection{Assumptions}  \label{assumptions}                                                                                                                                                                                                        

Our conclusions rest on rigorous theorems that apply to
compactified satisfying  certain 
conditions in addition to NEC:
\begin{itemize}
\item {\it GR condition:}
both the higher dimensional theory and the 4d theory are 
described by Einstein's 
theory of general relativity (GR), either exactly or with small 
corrections;
\item {\it Flatness condition:} the 4d theory is spatially flat;
\item {\it Boundedness condition:}
the extra dimensions are bounded;
\item {\it Metric condition:}
the metric of the higher dimensional theory is ${\cal R}$-flat (RF) 
or ${\cal R}$-flat up to a conformal factor (CRF):
\begin{equation} \label{metric}
ds^2  =  e^{2 \Omega} (-dt^2 + \bar{a}^2(t) d{\bf x}^2) + 
 g_{mn} dy^m 
dy^n,
\end{equation}
where the {\bf x} are the non-compact spatial dimensions; $y \equiv 
\{y^m\}$ 
are the extra 
dimensions;  $\bar{a}(t)$ is the usual FRW 
scale factor; and 
\begin{equation}\label{ceq}
g_{mn}(t,y) = e^{-2 \bar{\Omega}} \bar{g}_{mn}
\end{equation} 
where $\bar{g}_{mn}$ has Ricci 
(scalar) curvature ${\cal R}=0$, as evaluated in 
the compact dimensions. We do not require that
$\bar{g}_{mn}$ have zero 
Ricci tensor.  We call the metric
${\cal R}${\it-flat} (RF) 
if $\bar{\Omega} = \, const.$ and {\it conformally} 
${\cal R}${\it -flat} 
(CRF) if $\Omega(t,y)= \bar{\Omega}(t,y)$.  
  We will use 
indices $\{M, \, N\}$ to represent all $4+k$ dimensions, $\{\mu, \, 
\nu\}$ to represent the 
non-compact dimensions,  and $\{m, n\}$ to represent the extra 
dimensions.
\end{itemize}

These conditions are common to many models published in the 
literature. 
The {\it GR condition} dates back to the original Kaluza-Klein theory 
and underlies the idea of unified 
theories based on compactifying extra dimensions.   It is reasonable 
to expect corrections, 
such as higher derivative terms, in the higher and 4d effective 
theory.  So long as those 
are small, the theorems will apply with obvious caveats (as discussed 
in Sec.~\ref{SecV}).  The {\it spatial flatness condition} for the 4d 
theory 
is motivated by cosmological observations, {\it e.g.,} from WMAP 
\cite{Komatsu:2008hk}.  The {\it boundedness condition} on the extra 
dimensions 
is needed because the theorems rely on integrating fields 
and warp factors over the compact direction.  In particular, the 
boundedness condition insures that, if $\Omega$ is 
non-trivial and has
continuous first derivative, 
then the Laplacian $\Delta \Omega$ must be 
non-zero for some $y$; 
this fact is useful in some of the proofs. 

The {\it metric condition} is motivated by common 
constructions in the literature, especially string theory.   The 
original Kaluza-Klein model, 
the Randall-Sundrum models,
and Calabi-Yau based models are all RF; some useful theorems for this 
case were developed in Refs.~\cite{Wesley:2008fg, Wesley:2008de}.  
 Metrics of CRF type appear in warped 
 Calabi-Yau \cite{Giddings:2001yu} 
 and warped conifold 
\cite{Klebanov:2000hb} constructions 
(where they are 
sometimes referred to as conformally Calabi-Yau metrics).
Here we derive no-go theorems for 
both RF 
and
 CRF models.  Our constraints for RF and CRF are slightly different 
in terms of the number of extra dimensions and the moduli fields to 
which they apply.  However, the differences do not affect   our 
conclusions for practical cases relevant to string theory, M-theory, 
the Kaluza-Klein model, {\it etc.}, so we will only present the 
details 
for CRF models and ignore the fine distinctions.

\subsection{Detecting NEC violation}

In this subsection, we develop some basic relations that make it 
possible to detect easily when a higher dimensional theory is forced 
to violate the NEC.

To describe a spatially-flat FRW spacetime after dimensional 
reduction, the metric 
$g_{mn}(t,y)$ and warp function $\Omega(t,y)$ must be functions of 
time 
$t$ and the 
extra-dimensional coordinates $y^m$ only.   Following the convention 
in 
Ref.~\cite{Wesley:2008fg}, we parameterize the rate 
of change of 
$g_{mn}$  using quantities $\xi$ and $\sigma_{mn}$ defined by
\be
\frac{1}{2} \frac{d \, g_{mn}}{d \, t} = \frac{1}{k} \xi g_{mn} + 
\sigma_{mn}
\ee
where $g^{mn} \sigma_{mn} =0$ and where $\xi$ and  $\sigma$ are 
functions of time 
and the extra dimensions; this relation assumes the gauge
choice discussed in Ref.~\cite{Wesley:2008fg}.  

It is important to note that all discussions of the equation-of-state, 
the NEC, accelerated 
expansion, the energy-momentum tensor $T_{MN}$, and the pressure and 
density of any 
components always refer to Einstein frame quantities in either the 
higher dimensional or 
4d effective theory.   
The space-space 
components of the energy-momentum tensor  are block 
diagonal with a 
$3 \times 3$ block describing the energy-momentum in the 
three non-compact 
dimensions and a $k \times k$ block for the $k$ compact directions.  
The 0-0 component 
is the higher dimensional energy density $\rho$.  The 0-$m$
components 
are generally non-zero 
but will be of no special interest for our theorems.

Associated with the two blocks of space-space components of $T_{IJ}$ 
are two trace 
averages:
\begin{equation}
p_3 \equiv \frac{1}{3} \gamma_3^{\mu \nu} T_{\mu \nu}
 \; \; {\rm and} \; \;
p_k \equiv \frac{1}{k} \gamma_k^{mn} T_{mn},
\end{equation}
where $\gamma_{3,k}$ are respectively
the $3\times 3$ and $k \times k$ blocks of the higher
dimensional space-time metric.
Violating the NEC means that $T_{MN} n^M n^N <0$ for at least one null 
vector 
$n^M$ and at least one space-time point.  

Our approach in this paper is not to identify {\it all} cases where 
the NEC 
is violated, which can be complicated;
 rather we find simple methods for identifying a 
subset of cases where it must be violated.  For this purpose, the
following two lemmas, proven in 
Ref.~\cite{Wesley:2008fg},
are very useful: 

\vspace{0.1in}
\noindent
{\it Lemma 1:}  If $\rho+p_3$ or $\rho+p_k$ is less than zero for any 
space-time point, 
then the NEC is violated.  (Note that the converse is not true, 
$\rho+p_3 \ge 0$ and  $\rho+p_k \ge 0$ does not 
guarantee that the NEC is satisfied.)
\vspace{0.1in}

The second lemma utilizes the concept of $A$-averaged quantities 
introduced in 
Ref.~\cite{Wesley:2008fg}:
\begin{equation}
\langle Q \rangle_A \; = 
\left( \int Q \, e^{A\Omega} \sqrt{g} \; \text{d}^k y \right) {\Large 
/}
\left( \int e^{A\Omega} \sqrt{g} \; \text{d}^k y \right);
\end{equation}
that is, quantities averaged over the extra dimensions with weight 
factor $e^{A 
\Omega}$ where, for simplicity, we restrict ourselves to constant  
$A$.   
Using the fact that the weight function in the $A$-average is positive 
definite, a straightforward consequence is:

\vspace{0.1in}
\noindent
{\it Lemma 2:}  If $\langle \rho+p_3 \rangle_A <0$ or $\langle 
\rho+p_k \rangle_A \; 
<0$ for any $A$ and any $\{ t, \, {\bf x}\}$, then the NEC must be 
violated.
\vspace{0.1in}

As with the case of Lemma 1, this test is asymmetrical: finding 
an $A$-average less than zero 
proves NEC is violated, but finding a positive
average  is not 
sufficient to 
conclude NEC is satisfied.  

To illustrate the utility of $A$-averaging, we
introduce the CRF metric into the
the higher-dimensional Einstein equations, and then 
try to express terms dependent on $\bar{a}$ 
in terms 
of the 
 4d effective scale factor using the relation
$a(t) \equiv 
e^{\phi/2}\bar{a}(t)$, where 
\cite{Wesley:2008fg}: 
\begin{equation}\label{phieq}
e^{\phi} \equiv
\ell^{-k} \int e^{2\Omega} \sqrt{g} \; \text{d}^k y
\end{equation}
and  $\ell$ is the  
$4+k$-dimensional Planck length.
The 4d effective scale factor,  $a(t)$, obeys 
 the usual 4d Friedmann 
equations:
\begin{eqnarray}
\left( \frac{\dot{a}}{a} \right)^2 = \frac{1}{3} \rho_{4d} \\
\left(\frac{\dot{a}}{a}\right)^2 + 2 \frac{\ddot{a}}{a}= - p_{4d}
\end{eqnarray}
(henceforth, we use reduced Planck units,  
$8 \pi G_N=1$ in 4d; also, except where displayed explicitly, we 
choose $\ell=1$ in the $4+k$-dimensional theory).
Note that the 4d effective  energy density  $\rho_{4d}$ and pressure 
$p_{4d}$ are 
generally different from $\rho$ and $p_3$ in the higher dimensional 
theory if the warp factor $\Omega$ is non-trivial.  
Then, using the Einstein equations, we obtain
\begin{eqnarray}
\label{eq:QuinRhoPlusP3}
e^{-\phi}
\langle  e^{2\Omega} (\rho + p_3) \rangle_A   &= & (\rho_{4d} +p_{4d})
- \frac{k+2}{2k} \langle \xi \rangle_A^2 - \frac{k+2}{2k} \langle (\xi 
- \langle 
\xi\rangle_A)^2\rangle_A \;  - \langle \sigma^2 \rangle_A \; \\
\label{Astar}
e^{-\phi}
\langle e^{2\Omega} (\rho + p_k) \rangle_A \; & = & 
\frac{1}{2}\left( \rho_{4d} +3p_{4d} \right) + 2\left(\frac{A}{4} - 
1\right)\frac{k+2}{2k} \langle (\xi - \langle \xi\rangle_A)^2\rangle_A 
\; \notag \\
& & - \frac{k+2}{2k} \langle \xi \rangle_A^2 
  - \langle \sigma^2 \rangle_A \; \notag \\
& & + \left[ -5 + \frac{10}{k} + k + A\left(-3 + 
\frac{6}{k}\right)\right]
\langle e^{2\Omega} (\partial\Omega)^2 \rangle_A \; \notag \\
& & + \frac{k+2}{2k} \frac{1}{a^3} \,\frac{d}{d t} \left(a^3 \langle
\xi \rangle_A  \right)
\end{eqnarray}
$A$-averaging is a powerful tool because, with a judicious choice, one 
can insure that certain  
coefficients on the right hand side, the ones that depend explicitly 
on $A$,
 are non-positive. This opens a path for proving some of the no-go 
theorems below.  

This freedom is possible provided there is 
a range 
where
\begin{equation} \label{rangeA}
4 \ge A \ge \frac{ 10-5k+k^2}{3k-6} \equiv A_*,
\end{equation}
which is the case for $13 \ge k \ge 3$ (for CRF).  
Some theorems
below rely on choosing 
$A=2$; for this value to be within the range given in 
Eq.~(\ref{rangeA}), it is necessary that $8 \ge
k \ge 3$.  
(The corresponding ranges of $k$ in the RF case are given the Appendix.)
Since this includes 
the relevant string and M-theory models, we will implicitly assume 
this range of $k$ for CRF models for the remainder of this paper.  
(For $k=1$, the 
metric reduces to RF and similar theorems in Ref.~\cite{Wesley:2008fg} 
apply.)

The two relations in Eq.~(\ref{eq:QuinRhoPlusP3}) can be rewritten
\begin{eqnarray}
\label{essentials1}
e^{-\phi}
\langle  e^{2\Omega} (\rho + p_3) \rangle_A \;  &  = & 
\rho_{4d}(1  +w)
- \frac{k+2}{2k} \langle \xi \rangle_A \;^2 +               
{\rm non-positive \, terms \, for} \, {\it all} \, A
\\  \label{essentials2}
e^{-\phi}
\langle e^{2\Omega} (\rho + p_k) \rangle_A \; & = & 
\frac{1}{2}\rho_{4d}(1+3 w) + \frac{k+2}{2k} \frac{1}{a^3} 
\,\frac{d}{d t} 
\left(a^3 \langle \xi \rangle_A  \right) \notag \\
& & +{\rm non-positive \, terms \, for}\, {\it some} \, A, 
\end{eqnarray}
where the values of $A$ that make the 
last term non-positive  are those that are 
in the range in 
Eq.~(\ref{rangeA}).  
Henceforth, unless stated otherwise, we always choose
$A$ to be in that range.
Recall that
$w$ represents the ratio of the 
total 4d effective 
pressure $p_{4d}$ to the total 4d effective energy density 
$\rho_{4d}$.  In Appendix I, we provide the coefficients 
of the last term in Eqs.~(\ref{essentials1}) 
and~(\ref{essentials2})
for the case where the moduli are frozen $\xi=0$.

On the left hand side of Eqs.~(\ref{essentials1}) 
and~(\ref{essentials2}), 
both $\phi$ and 
$\langle \ldots \rangle_A$ depend 
on the warp factor, $\Omega$,  but the
combination is invariant under shifts $\Omega \rightarrow \Omega +C$, where $C$ is a constant.  
Furthermore, the combination tends to have 
a weak dependence on $\Omega$.
 For example, if $\rho+ p_k$ is 
homogeneous in $\{y^m\}$, the left hand side reduces to
$ K (\rho+p_k)$,  where the dimensionless
coefficient $K$ is  not very 
sensitive to $\Omega$ or $A$; in particular,
$K = \ell^k \,I(A+2)/I(A)I(2)$ where
\begin{equation}
I(\bar{A}) \equiv
\int e^{\bar{A} \Omega} \sqrt{g} \; \text{d}^k y.
\end{equation}
In this notation, the $k$-dimensional 
volume of the compact space  is 
$V_k=I(0)$; then, $K$ is equal to $\ell^k/V_k$, a coefficient which is strictly less than unity.  
Similarly, if $\rho+p_k$ is smooth and $\Omega$ has a sharp maximum on some subspace of dimension $m$ and volume $v_m$, 
then 
the left hand side of Eq.~(\ref{essentials2})  is 
${\cal O}(1)(\ell^m/v_m)(\rho+p_k)_{max}$, where $(\rho+p_k)_{max}$ is the value of $\rho+p_k$ 
evaluated on the subspace where $\Omega$ is maximal. 
We will use this example in Sec.~\ref{SecViolate}.

\section{No-go Theorems for 
Models that Satisfy NEC}  \label{SecIII}

The lemmas of the previous subsection can be used to prove 
that compactified theories satisfying NEC and meeting the other 
assumptions given at the beginning of Sec.~(\ref{SecII})
are 
incompatible with inflation and the simplest
dark energy models consistent with observations.
The theories include the original Kaluza-Klein model and many string 
theories. The Randall-Sundrum II model\cite{RS2},
with a single brane, also satisfies NEC;
 formally,
it does not satisfy the boundedness condition,
but, because the 
warp factor is well-behaved at infinite distances from the brane, we 
conjecture that the same theorems apply.

As a first step, we show that $w$ must be strictly greater 
than -1. The argument is simple.  If $w=-1$, the first term in 
Eq.~(\ref{essentials1}) is precisely zero and the 
second two are non-positive.   Consequently, NEC can only be 
satisfied if the last two terms are precisely zero as well.  However, 
in 
Eq.~(\ref{essentials2}), the first term is strictly negative and the 
last term is non-positive.  
Hence, the middle term must be positive for this equation to to 
satisfy the NEC; but this 
requires $\xi$ and/or its time-derivative to be non-zero.  But this is 
incompatible with having 
the middle term in Eq.~(\ref{essentials2}) be zero.  Hence, one or 
both 
equations must violate the 
NEC if $w=-1$.  

An immediate consequence is a first dark energy 
no-go theorem.  (Theorems labeled IA, IB, {\it etc.} refer to models 
obeying NEC and models labeled IIA,IIB, {\it etc.} refer to models 
that violate NEC.) 

\vspace{0.1in}
\noindent
{\it Dark Energy No-go Theorem IA:}  $\Lambda$CDM  (the current 
concordance model in cosmology) is incompatible with
compactified 
models\cite{note} satisfying the NEC.
\vspace{0.1in}

A pure de Sitter universe is obviously ruled out by the argument 
above.
Also, $w <-1$ is forbidden by the assumption that the 4d effective 
theory obeys the NEC.  It is further apparent that  $w>-1$ but close 
to -1 is subject to the same problems. Consequently, a $\Lambda$CDM  
universe, with a mixture of matter and positive cosmological constant 
that approaches $w=-1$ in the future, is also ruled out.

This point can be made more forcefully and precisely.   
Depending on the number of extra dimensions 
and whether the metric is RF 
or CRF, there exists a $w_{transient}$ between $-1/3$ and $-1$ 
such that $w$ can only remain in the interval
$ (-1, \; w_{transient})$  for a few e-folds.
The condition $w< w_{transient}$ cannot be maintained indefinitely 
because it requires  either NEC 
violation of Eq.~(\ref{essentials2}), if the average $\xi$ and $d 
\xi/dt$ are kept small or 
negative; or NEC violation of Eq.~(\ref{essentials1}), if $\xi$ is 
made large and positive 
enough to avoid NEC violation in Eq.~(\ref{essentials2}).  In 
principle, it is possible 
to satisfy NEC for both relations if
 $\xi$ is near zero and 
$d \xi/dt$ is large and positive, but this can only be maintained for 
a brief period.

How brief is brief?  In order for the right hand side of 
Eq.~(\ref{essentials2}) to remain positive, it 
is necessary that 
\begin{equation} \label{pos}
\frac{k+2}{2k} \frac{1}{a^3} \,\frac{d}{d t} \left(a^3 \langle \xi 
\rangle_A \;\right)  > - \frac{1}{2}\rho_{4d}(1+3 w).
\end{equation} 
The right hand side is positive for $w< w_{transient}$ and has 
magnitude  ${\cal O}( \rho_{4d})= {\cal O}(H^2)$, where $H \equiv 
\dot{a}/a$ is the Hubble parameter.  Hence, $H^{-2} d \langle \xi 
\rangle_A \;/dt  = {\cal O}(1)$.  Now suppose $ \langle \xi \rangle_A  
$ begins small so that Eq.~(\ref{essentials1}) is satisfied initially.  
Integrating over a Hubble time, we find that $ \langle \xi \rangle_A  
$, grows until 
\begin{equation} \label{xi_cond}
\langle \xi \rangle_A \;/H  = {\cal O}(1)
\end{equation}
at which point Eq.~(\ref{essentials1}) violates NEC.
In other words, the brief period during which $\xi$ remains 
small  cannot last more than 
a few Hubble times. 

To reach $w < w_{transient}$ in the first place, it must be that 
$w_{\rm{DE}}$  is less than $w_{transient}$. 
But, then, the  only way to avoid violating NEC is for $w$ to
increase above $w_{transient}$ after a few e-folds, which is only 
possible if $w_{\rm{DE}}$ itself 
increases 
above $w_{transient}$ after a few e-folds (which we will take to be three 
e-folds, for the purposes of this paper).
 
\begin{figure} 
\epsfxsize=5 in \centerline{\epsfbox{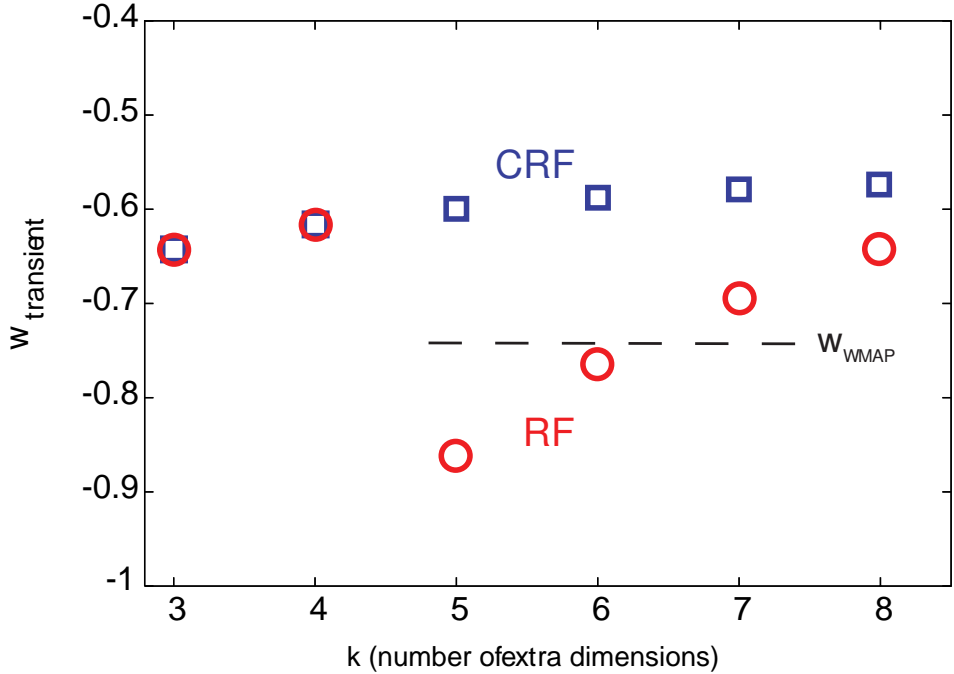}}
 \caption{ Plot of $w_{transient}$ vs. the number of extra dimensions 
$k$ for extra-dimensional models based on
 RF (circle) and CRF (square) metrics.  The dashed horizontal segment 
represents the current value of $w$ according to WMAP.
	    } \label{figtrans}
	      \end{figure}

A plot of $w_{transient}$ as a function of the number of extra 
dimensions is given in Fig.~\ref{figtrans}; note that $w_{transient}$ 
is 
substantially greater than $-1$ for 
the cases of greatest interest, such as string theory ($k=6$)
or M-theory ($k=7$).   
(See Ref.~\cite{OtherPaper} for a more detailed quantitative 
discussion).

Two additional no-go theorems follow from this analysis:

\vspace{0.1in}
\noindent
{\it Dark Energy No-go Theorem IB:}  Dark energy models with constant 
$w_{\rm{DE}}$ less than $w_{transient}$ or time-varying $w_{\rm{DE}}$ 
whose 
value 
remain less than $w_{transient}$ for a continuous period lasting more 
than a 
few Hubble times are incompatible with
compactified models\cite{note} satisfying 
the NEC.
\vspace{0.1in}

This theorem rules out a wide spectrum of dark energy models, 
including 
a 
range which is currently allowed observationally and that the JDEM 
mission is designed to explore \cite{Albrecht:2006um}.  Conversely, if 
JDEM 
indicates $w_{\rm{DE}} < w_{transient}$ and constant, this would
rule out this entire class 
of compactified models. 

\vspace{0.1in}
\noindent
{\it Inflationary No-go Theorem IA:}  Inflationary models consistent 
with observations are incompatible with 
compactified models\cite{note} satisfying 
the NEC. 
\vspace{0.1in} 

Inflationary cosmology requires a period of 40 or more e-folds of 
accelerated expansion 
with $w \approx -1$ to within a few percent in order to smooth and 
flatten the universe 
and to obtain a scalar spectral index within current observational 
bounds \cite{Komatsu:2008hk}.  
For the compactified models considered here,
this value of $w$
is far below $w_{transient}$ and cannot be maintained for 
more than a few e-folds -- certainly not for 40 e-folds.  

\vspace{0.1in}
\noindent
{\it Inflationary Corollary:} Compactified models\cite{note} 
satisfying the 
NEC 
 are 
counterexamples to the common assertion that 
inflation with nearly scale-invariant spectra are 
an inevitable 
consequence given chaotic or 
generic initial conditions after the big bang.
\vspace{0.1in}

The common lore is that, after the big bang,
the universe is chaotic but there are always
rare patches of space that are smooth enough 
and have the right 
 conditions to initiate inflation (assuming an 
 inflaton with a sufficiently flat potential); 
 and these patches  soon 
dominate the volume of the universe.  
For the 
entire class of theories considered here, though,
no patches of space 
undergo inflation that is slow and long-lasting enough to produce a 
spectral tilt anywhere near the observational bounds.  The problem 
is not finding a scalar field with a sufficiently flat potential, 
because none of the compactified models
explicitly forbids that. Rather, 
the problem is that accelerated 
expansion induces a rapid variation of  $\xi$ which, 
in the 4d effective theory in Einstein frame, 
 appears as a time-varying field whose 
kinetic energy increases the overall equation of 
state $w$  and prevents
inflation from continuing 
long enough.

Additionally, for  all models (RF or CRF, and for 
any $k$), $\langle \xi \rangle_A = 
\dot{G}_N/G_N $ for  $A=2$,  
if the 
theory is expressed in Jordan-Brans-Dicke 
(JBD) frame. Although we keep to Einstein frame in this paper 
generally, it useful to express $\xi$ in 
terms of $\dot{G_N}$  for the purpose of comparison to observational 
constraints on $\dot{G_N}$ which implicitly assume 
JBD frame.  
For RF models, $A=2$ lies outside the
corresponding range of $A$ (see Appendix A) and so theorems below about changing $G_N$ should be re-expressed as conditions on changing $ \xi$ or, equivalently, changing size of the extra dimensions.
For CRF models,
$A=2$ lies in the range in Eq.~(\ref{rangeA})
when $8 \ge k \ge 3$.
Converted to JBD frame, we could restate our conclusion for CRF models
as follows:
accelerated 
expansion induces a rapid variation of the gravitational constant,  $ 
\dot{G}_N/G_N H= 
{\cal O} (1)$.
. 

Returning now to Eq.~(\ref{pos}), we note that 
it requires
that, regardless of the value of $A$, 
$d \, 
 \langle \xi \rangle_A/dt =  {\cal O}(H^2)$  be positive not only for 
$w<w_{transient}$,
 but also for
$w<-1/3$. From this emerges:

\vspace{0.1in}
\noindent
{\it Dark Energy No-go Theorem IC:} All dark energy models are 
incompatible with 
compactified models\cite{note} satisfying the NEC
 if the moduli
fields are frozen
 (or, specifically, $G_N$ is constant, in the case of CRF models).
\vspace{0.1in}

\noindent
This follows trivially because any form of dark energy requires $w$ 
reach a value 
less than -1/3, and, as we just argued, $\xi$ must vary with 
time whenever $w< -1/3$ if NEC is satisfied.

As a practical matter,
the current value of $w \approx -0.74$ is already less than 
$w_{transient}$ for $k=7$ dimensions (e.g., M-theory).  
In this case, both 
Dark Energy no-go Theorem IC and
IB apply, but Theorem IB is more stringent.  
Theorem IB says that both 
$w_{\rm{DE}}$ and $G_N$ must vary with time and at high rates. 
One might wonder: Is it already 
possible to rule out all RF and CRF
compactified models\cite{note} satisfying the NEC 
based on current observations?
In Ref.~\cite{OtherPaper}, we show that the answer is no; there 
remains a small window  in the parameter space 
$\{ w_{\rm{DE}}, \, dw_{\rm{DE}}/dt, \, \dot{G}_N/G_N \}$ 
consistent with all current observations.  However, anticipated 
observations 
will be able to
check this remaining window to determine if this class of theories is 
empirically ruled out or not.

\section{No-go Theorems for  
Models that Violate NEC} \label{SecIV}

In this section, we continue to consider compactified models 
satisfying the GR, flatness, boundedness, and metric conditions 
assumed in Sec. II.  The difference  is that, before, we 
only considered models that satisfy the NEC, in which case we showed 
that 
moduli fields $\xi$ (and $G_N$) must vary with 
time at a fast rate barely compatible with 
current observational constraints and potentially  ruled out by 
near-future 
observations.  So now  we consider models that violate NEC but with 
 fixed (or very slowly varying) moduli.    
Theories of this type include the Randall-Sundrum I 
model\cite{RS1}, because it includes a negative tension brane,
and some models that arise 
in flux compactifications 
of Calabi-Yau 
manifolds when NEC-violating components, such as 
orientifold-planes or 
Casimir energy, are 
introduced.  

If the only requirement for incorporating accelerated expansion were 
NEC violation, then it would suffice 
if  $\rho+ p_3 <0$ or $\rho+p_k<0$ at any one space-time point.  
However, we now will present a set of no-go theorems that show that 
cosmic 
acceleration imposes a host of stringent conditions on 
the spatial 
distribution and temporal variation of the NEC-violating 
elements.  The no-go theorems in this section 
are qualitatively the 
same for dark energy and inflation because the theorems only rely on 
the fact that the 
universe must evolve from 
$w>-1/3$ to  $w<-1/3$ or vice versa, which is required both for 
inflation and dark energy cosmology.  
Recall that $w$ refers to the ratio of 
the total pressure ($p_{4d}$) to the energy 
density ($\rho_{4d}$) in the 4d effective theory.   

\vspace{0.1in}
\noindent
{\it Inflationary/Dark Energy No-go Theorem IIA:} Inflation and dark 
energy
are incompatible with 
compactified models\cite{note} (with fixed moduli) if 
the NEC is satisfied in the compact dimensions ({\it i.e.,} $\rho 
+ p_k \ge 0$ for all $t$ and $y_m$) --- whether or not NEC is violated in 
the non-compact directions.
\vspace{0.1in}

The first step in the proof is to note that, since $G_N$ (and other 
moduli) are assumed to be fixed, the middle term in the expression for 
$e^{-\phi}\langle e^{2 \Omega} (\rho+ p_k )\rangle_A$ in Eq.~(\ref{essentials2}) 
is zero. In this case, the relations in the appendix apply.  We can 
use 
the freedom to choose $A$ in our 
$A$-averaging so that the third term in Eq.~(\ref{essentials2}) is 
zero; this corresponds to  $A=A_*$ in Eq.~(\ref{rangeA}).  
  That leaves only the 
the first term, 
proportional to $1+ 3 w$, which is positive for $w>-1/3$ and negative 
for $w<-1/3$.   Hence, whenever the 
universe is accelerating ($w< -1/3$), NEC violation must occur in the 
compact dimensions.   (It may or may not occur in the non-compact 
dimensions as well.)

\vspace{0.1in}
\noindent
{\it Inflationary/Dark Energy No-go Theorem IIB:} Inflation and dark 
energy
are incompatible with 
compactified models\cite{note} (with fixed moduli) 
for which
the net NEC violation ($\rho+p_k$) is time-independent.
\vspace{0.1in}

This theorem relies on the fact that 
both inflation and dark energy models have a transition from 
 phases with $w> -1/3$ to phases with $w <-1/3$.   (This proof does 
not apply to a pure de Sitter phase where $w$ is always equal to -1.)
Since $e^{-\phi}\langle e^{2 \Omega}(\rho+ p_k) \rangle_{A_*}$  
is proportional 
to $3 w+1$, which switches sign 
as $w$ evolves past $w =-1/3$, the 
NEC violation (summing over all energy density and
pressure contributions) in the compact 
direction must be time-dependent.  In the case of inflation, there is 
also a transition in which $w$ changes from less than $-1/3$ to 
greater than $-1/3$. This leads to an important corollary:

\vspace{0.1in}
\noindent
{\it Inflationary Corollary:} 
Inflationary cosmology is only 
compatible with compactified theories\cite{note}  that include an NEC 
violating component in the compact dimensions whose magnitude is of 
order the vacuum density (that is,
$e^{-\phi} \langle e^{2 \Omega} (\rho+ p_k) \rangle_{A_*} \sim \rho_{4d}$); 
such  
that $\langle\rho+ p_k\rangle_{A_*}$ switches from positive to 
negative 
when inflation begins {\it and} 
 switches back when inflation is complete.
\vspace{0.1in}

The corollary means that the requirements usually associated with 
inflation --- a scalar field with a flat potential, stringent 
conditions on slow-roll parameters, a reheating mechanism, {\it etc.} 
---  are not sufficient to have inflation in compactified theories 
since they do not produce or annihilate NEC violations.  Furthermore, 
the magnitude of the NEC violation is one hundred orders of magnitude 
larger than what is required to support a dark energy phase; so the 
source of NEC violation must be different from whatever is used to 
produce the current vacuum state.  Finally, after inflation is over, 
$e^{-\phi} \langle e^{2 \Omega} \rho+ p_k\rangle_{A_*}$
 must switch sign again, so the 
reheating in the non-compact dimensions must somehow have 
back-reaction that changes the NEC violation in 
the compact directions by a 
hundred orders of magnitude.

There is more to be said.
Theorem IIA imposes the condition that NEC is violated in the compact 
dimensions. The next no-go theorems constrain the spatial distribution 
of the NEC-violating elements within those compact directions.

\vspace{0.1in}
\noindent
{\it Inflationary/Dark Energy No-go Theorem IIC:} Inflation and dark 
energy
are incompatible with 
compactified models\cite{note} with fixed moduli if the warp factor 
$\Omega(t,y)$ is non-trivial and has
continuous first derivative and if any of the 
following quantities is homogeneous in $y$:
\begin{enumerate}
\item $\rho + p_3$; 
\item $x \rho + p_k$ for RF metric, for any $(1/2) (1-3w) > x > 4(k-
1)/3k$;
\item $\rho$ for CRF metric for $k> 4$;
\item $2 \rho + p_k$ for CRF metric for $k> 3$ and $w>-1$;
\end{enumerate} 
\vspace{0.1in}

The first condition follows straightforwardly from 
Eqs.~(\ref{usefulrf}) and~(\ref{usefulcrf}), 
which show that $\rho+ p_3 
= e^{\phi-2 \Omega} (\rho_{4d} + p_{4d})$.  This expression must be 
inhomogeneous because $\Omega$ is $y$-{\it dependent} (by assumption) and 
 the 4d effective energy density $\rho_{4d}$ and pressure $p_{4d}$ are 
$y$-{\it independent} (by definition).

The remaining conditions are proven by using Eq.~(\ref{usefulcrf}) in 
the appendix to express 
each of the linear combinations of $\rho$ and 
$p_k$ in the list above as:
\begin{equation} \label{rhoeq}
C\,  \Delta \Omega  + D (\partial \Omega)^2 + E e^{-2 \Omega} 
\rho_{4d},
\end{equation}
where $C$ and $E$ are have the same sign.  For example, consider the 
case where $C$ and $E$ are positive.   If $\Omega$ is non-trivial and 
has 
continuous first derivative
and if the compact dimensions are bounded, then $\Omega$ 
must have a non-zero maximum and minimum on the compact manifold.  At 
the maximum, we have that $\partial \Omega=0$ (so the middle term is 
zero), $ \Delta \Omega<0$ and $e^{-2 \Omega}$ is minimal; similarly, 
at the minimum, the middle term is also zero but $ \Delta \Omega>0$ 
and $e^{-2 \Omega}$ is maximal.  Hence, for positive $C$ and $E$, 
both terms in Eq.~(\ref{rhoeq}) are smaller for maximal $\Omega$ 
compared to their values for minimal $\Omega$; the sum cannot be a 
homogeneous function of $y$.  (A similar argument 
applies if $C$ and 
$E$ or both negative.)
  
For the RF case, a similar argument can be used to
show that $x \rho +p_k$ must be 
inhomogeneous for a continuum of set of choices $(1/2) (1-3w) > x> 4 
(k-1)/3k$.   Note that there exists a non-zero range of $x$ provided 
$w< - 5/9 + (8/9k)$, which includes all $w<-5/9$.  Since all 
observationally acceptable dark energy and inflation models must pass 
through phases where $w<-5/9$, 
these models require $x \rho +p_k$ be inhomogeneous for a finite range 
of $x$.  A similar argument holds for the CRF case, but here we have, 
for 
simplicity, limited ourselves to two linear combinations: $\rho$ alone 
and $2 \rho +p_k$, which must both be inhomogeneous 
for all $w>-1$.  

We have made no attempt to be exhaustive here because these examples 
suffice to make the point that the energy density and pressure must 
have non-trivial distributions across the extra dimension to satisfy 
the higher dimensional Einstein equations.  Further constraints are 
given 
by  the 
following no-go theorems that rely on somewhat different methods of 
proof.

\vspace{0.1in}
\noindent
{\it Inflationary/Dark Energy No-go Theorem IID:} Inflation and dark 
energy
are incompatible with 
compactified models\cite{note} with fixed moduli if the warp factor 
$\Omega(t,y)$ is non-trivial if $\rho + p_k$ is homogeneous.
\vspace{0.1in}

Note that this linear combination is the indicator of NEC violation, 
so this no-go theorem says that that the degree of NEC violation must 
itself be inhomogeneously distributed in the compact dimensions.  To 
prove this result, it suffices to restrict ourselves to showing that 
$\rho+ p_k$ is inhomogeneous for $w=-1/3$ since both dark energy and 
inflation models must pass through this value of $w$.   For $w=-1/3$,
the last term in $\rho +  p_k$ in Eq.~(\ref{usefulrf}) (for RF) and 
Eq.~(\ref{usefulcrf}) (for CRF) in the Appendix is zero.  
Using Lemma~A1 in the Appendix, the remaining terms can be rewritten 
as $\Gamma e^{-\gamma \Omega} \Delta e^{\gamma \Omega}$ where $\gamma$ 
and $\Gamma$ are positive.  For non-trivial $\Omega$, $\Delta \Omega$ 
must be non-zero and have different signs at the maximum and minimum 
of $\Omega$ on the compact manifold.  Hence, $\rho+p_k$ must be 
inhomogeneous.

\vspace{0.1in}
\noindent
{\it Inflationary/Dark Energy No-go Theorem IIE:} Inflation and dark 
energy
are incompatible with 
compactified models\cite{note} with fixed moduli 
if $w_k(A_*) \equiv \langle p_k \rangle_{A_*}/\langle \rho_k 
\rangle_{A_*} > -1$ for $\langle \rho\rangle_{A_*}>0$ or if $w_k(A_*) 
\equiv \langle p_k\rangle_{A_*}/\langle\rho_k\rangle_{A_*} < -1$ for 
$\langle\rho\rangle_{A_*}<0$ .
\vspace{0.1in}

We will present the proof before explaining its significance: Let us 
first consider the case where $\langle \rho \rangle_{A_*} > 0$.
Based on 
Eqs.~(\ref{wRF}) and~(\ref{wCRF}) in the Appendix, we can express 
$w_k(A)$ as:
\begin{equation} \label{ratkey}
w_k(A) = \frac{ g(A) \langle
(\partial \Omega)^2\rangle_A + \frac{3 w-1}{2}X} { f(A) 
\langle (\partial \Omega)^2 \rangle_A + X},
\end{equation}
where $X = \langle e^{\phi-2 \Omega} \rho_{4d} \rangle_A >0$. (Recall that 
$\rho_{4d}>0$ 
in  inflation and dark energy models.)
Recall that the denominator is 
$<\rho_A$, the $A$-averaged energy density.  For $A = A_*$ (as given 
in 
Eq.~(\ref{rangeA})), $f(A_*)/g(A_*) = -1$.  For $w <-1/3$, as 
required for inflation or dark energy models, the coefficient of $X$ 
in the numerator is less than -1. Straightforward algebra then shows 
that $w_k(A_*)$ is strictly less than $-1$.  (A similar argument
can be used to show $w_k(A_*)$  is strictly greater than 
$-1$ if   
$\langle \rho \rangle_A < 0$.)

The quantity $w_k$ is the ratio of the volume-averaged pressure to 
volume-averaged energy  density with positive definite weight $e^{-2 
\Omega}$.  To have NEC violation in the compact dimensions, as 
required by Theorem IIA, it suffices that $p_k/\rho <-1$ for $\rho >0$ 
at a single point; or  $p_k/\rho > -1$ for $\rho < 0$ at a single 
point. Here we have shown that the ratio volume weighted averages must 
satisfy these inequalities, generally a much stronger condition. 

This no-go theorem is 
useful because it shows that simply violating the NEC is not enough; 
one must be deeply within the NEC-violating regime.
For example, for constant warp factor 
$\Omega$, $w_k(A) = (3 w-1)/2$ (independent 
of $A$), which approaches $-2$ as $w \rightarrow -1$.  This value is 
far 
below the minimal value needed to violate the NEC; 
{\it e.g.,}
inconsistent 
with simply Casimir energy or a single orientifold-plane
as the source of 
NEC violation. 

There are some other curiosities. For example, for constant warp 
factor
$\Omega$, radiation alone exerts positive pressure in the non-compact 
dimensions, but must exert zero pressure in the compact dimensions; 
and matter exerts no pressure in the non-compact dimensions, but must 
exert negative pressure int the compact dimensions.

\section{Constraints on Models violating the GR or metric 
conditions}\label{SecViolate} 

Formally, the theorems derived here apply strictly to models in which the 
higher dimensional theory satisfies Einstein's 
equations and is 
described by an RF or CRF metric. However, the theorems provide useful insights for some models that violate one or both conditions. For example, some string inflation models
satisfy the GR conditions perturbatively but violate them non-perturbatively
\cite{Kachru:2003aw,Giddings:2001yu,Baumann:2008kq}. 
One might inquire whether these models evade the no-go theorems derived in this paper.  
Absent an explicit expression for the non-perturbative interactions, a quantitatively precise answer
cannot be reached.  Nevertheless, qualitatively, it is clear that the no-go theorems may only be evaded if the violations are large and time-dependent.  

For example, if the 
violations can be 
expressed as additions to the right-hand-side of 
Eqs.~(\ref{essentials1}) and~Eq.~(\ref{essentials2}), 
then,  
these modifications
 have to balance the equations by satisfying 
similar time-variation conditions 
as required for the NEC-violating components in the proofs of the 
no-go theorems.   That is, there must be some 
sort of back-reaction
in the compact directions in either case.
By the argument given below Eq.~(\ref{essentials2}),
the modifications 
to  $e^{-\phi}\langle e^{2 \Omega}(\rho+ p_3) \rangle_A$ 
and $e^{-\phi}\langle e^{2 \Omega}(\rho+ p_k) \rangle_A$ must be of order 
$\rho_{4d}$, and they must change by an  amount ${\cal O}(1) \rho_{4d}$
 whenever the universe 
switches from accelerating to decelerating (or vice versa) in order to 
change the sign of $e^{-\phi} \langle 
e^{2 \Omega} \rho + p_k \rangle_{A_*}$ 
(as required by the
kind of argument presented for Theorem IIB).  

What makes the back-reaction
problematic is that, phenomenologically,
the change from acceleration 
to deceleration (or vice versa) 
in the 4d effective theory is supposed to be due entirely 
to the production of matter and radiation (in the case of inflation) 
or red-shifting of matter energy density (at the onset of dark energy
domination) that acts in the non-compact dimensions; 
so it would seem that
any back-reaction in the compact dimensions required to satisfy Eq.~(\ref{essentials2}) had better turn out to be quantitatively small enough to have a negligible effect
on the 4d effective theory.  If the 
effect of back-reaction on the 4d effective theory is not negligible, it will alter the course of 
accelerated expansion in undesirable ways, such as 
shortening or eliminating the acceleration phase, 
as was shown to be the case for models that satisfy the 
GR and metric conditions.  In the case of inflation, 
even if the back-reaction does not prevent inflation, 
it may change the transition from inflation to reheating 
and, thereby, the predictions.  

In fact, in certain flux compactifications in string theory,
there is an argument to suggest that the back-reaction will have a very large effect. These models invoke 
orientifold-planes (extended objects with negative tension) that serve as sources of the NEC violation necessary to stabilize a true vacuum with positive cosmological constant.\cite{Kachru:2003aw,Giddings:2001yu}  
Averaged over the bulk volume, the large negative tension of the orientifold-planes 
is nearly canceled by large positive density contributions, such as branes. There can also be positive density contributions in the throat.
However, several of 
the no-go theorems entail the $A_*$-average
of $\rho+p_k $ where $A_* \ge 1$. For example,
Theorem IIB 
 requires that this average
 switch sign and change by an amount of order $\rho_{4d}$ when the universe transitions from acceleration to deceleration (or vice versa). 
 Because the  $A_*$-average 
over the compact volume weights contributions to $ \rho+p_k $ by a factor of
$e^{A_* \Omega}$,  
 contributions  from regions in the compact volume where $\Omega$ is maximal will be strongly weighted compared to regions where $\Omega$ is small. 
In the case of orientifold-planes, singular surfaces near 
which $G_{00} <0$ and  $(\partial \Omega)^2$ approaches zero, 
Eq.~(\ref{usefulcrf}) implies $\Delta \Omega<0$; hence, orientifold-planes are (local or global) maxima of $\Omega$ and tend to be strongly weighted in the $A_*$-average. 

Consider, for example, 
a setup where $\Omega$ is maximal  along the 
orientifold-planes  which have some constant $(\rho+p_k)_{neg}<0$ in a volume of dimension $m<k$ and volume $v_m$; further suppose that
$\Omega_{pos}$ is somewhat smaller but nearly uniform over the rest of the
bulk where there is some average stress-energy $(\rho+p_k)_{pos}> 0$ that nearly balances the orientifold-plane component; finally, as in the case of $d$-brane inflation, suppose there is some positive 
$(\rho+p_k)_{throat}>0$ contribution in the throat.  The $A_*$-weighted 
combination of these components is then:
\begin{equation} \label{neweq}
\frac{\ell^m}{v_m} (\rho+p_k)_{neg} + 
\frac{\ell^{2 m-k} V_k}{v_m^2} e^{-(A_*+2)\Delta \Omega_{bulk}}(\rho+p_k)_{pos} 
+\frac{\ell^{2 m-k} V_k}{v_m^2} e^{-(A_*+2)\Delta \Omega_{throat}}(\rho+p_k)_{throat}
\end{equation}
where $\Delta \Omega_{bulk} = \Omega_{neg}-\Omega_{pos} >0$ and
 $\Delta \Omega_{throat} = \Omega_{neg}-\Omega_{throat} >0$.
This sum is supposed to switch
 from an amount of order $-\rho_{4d}$ to $+\rho_{4d}$ at the end of inflation (or the reverse at the onset of dark energy domination).  Because of the $A_*$ weights, the exponentially dominant contribution to  Eq.~(\ref{neweq}) is the due to the orientifold-plane, which contributes an amount 
$(l^m/v_m) (\rho+p_k)_{neg}$, that 
is exponentially enhanced compared to the positive energy density contributions in the bulk or in the throat because, by
assumption,
the warp factor $\Omega$ is much larger  
near orientifold-planes.  
In order for $e^{-\phi} \langle e^{2 \Omega} 
\rho+p_k\rangle_{A_*}$
 to switch sign when the universe changes from accelerating to decelerating (or vice versa), the back-reaction in the bulk must 
either change the contribution of the
orientifold-planes by an amount of order $\rho_{4d}$, which seems unlikely 
given their topological character; or the back-reaction must change
the positive energy density  components by an exponentially larger amount. 
In the latter case especially, the effect of the 
back-reaction on the effective 4d theory is likely to be 
overwhelmingly
large.  (One could switch the scenario so that $\Omega$ is maximal in the bulk 
positive $(\rho+p_k)$ regions and smaller on the orientifolds; even so, the only way to change the sign on the left-hand side of Eq.~(\ref{essentials2}) is to have a back-reaction in 
which  some energy components change by an amount of at least  $\rho_{4d}$; and, in most cases, by an amount exponentially greater amount.)  It is, therefore, essential to track the effect of this back-reaction on the 4d effective theory (where the leading contribution is supposed to be of order $\rho_{4d}$) to be sure the cosmological scenario is not spoiled.
As of this writing, though, the back-reactions during the transition from inflation to reheating and from matter domination to dark energy domination are not well understood:
In particular, they have not been included in string inflation calculations
and predictions or in discussions of stringy dark energy models.

We note that our analysis has been restricted to the case of RF or CRF metrics which are Ricci flat or conformally Ricci flat and that we have ignored non-perturbative corrections to GR.  However, a similar argument applies if they are included. They can be viewed as amendments to the right-hand side of Eq.~(\ref{essentials2}); then, by the same reasoning, they must change by an amount of order $\rho_{4d}$ to balance the equation.  So, as in the case above, one must be concerned about the effect of their back-reaction in the 4d effective theory.

\section{Conclusions} \label{SecV}

The essence of this paper is that cosmic acceleration is surprisingly 
difficult to incorporate in compactified models.  The problem arises 
in trying to satisfy simultaneously the 4d and higher dimensional 
Einstein equations.  Both must be satisfied for any equation-of-state, 
but we have shown that,  for the metrics assumed in this 
paper,
 this requires increasingly exotic conditions as 
the universe goes from decelerated to accelerated expansion or, 
equivalently, as $w$ 
decreases below $-1/3$.  For dark energy models,
 either moduli fields 
(including $G_N$) have to change with time at a rate that is nearly 
ruled out (and may soon be excluded observationally 
altogether\cite{OtherPaper}) 
or NEC must be violated. 
For inflation, only the second option remains viable.

If the NEC is
violated, it must be violated in the compact dimensions; it must be 
violated strongly ($w_k$ significantly below the minimally 
requisite value for NEC violation); and the violation in the compact 
dimensions must vary with time in a manner that precisely tracks the 
equation-of-state in the 4d effective theory. For example, in 
realistic cosmological models, there are 
known 
matter and radiation components (baryons and photons, for example) 
that contribute to the energy and density of the 4d effective theory 
but are not normally related to NEC violation.   
Nevertheless, the no-go 
theorems say that the magnitude of NEC violation must vary with time
in sync with how the conventional matter and radiation 
energy density and pressure evolve.

Satisfying these equations for $\Lambda$CDM is difficult, but 
satisfying them for inflation is even harder.  A period of 
inflation 
with $w$ within a 
few percent of $-1$ (as required to meet the observational constraints 
on the spectral tilt) must be sustained for at least 40 e-folds to 
resolve the flatness and homogeneity problems;
this requirement restricts us to the 
case that the NEC is violated, according to Inflationary 
Theorem IA.   
The magnitude of the NEC violation is proportional to $\rho_{4d}$ 
according to 
Eq.~(\ref{essentials2}), which  is roughly $10^{100}$ times greater 
during the 
inflationary epoch than during the present dark energy dominated 
epoch. Hence, the source of 
NEC violation for inflation must be different and $10^{100}$ stronger.  
Also, identifying a scalar 
inflaton field with a flat potential or  branes and antibranes 
approaching one another  in some 
warped throat does not suffice because they do not violate NEC, 
either.  For example, as a hypothetical,
imagine that  a D3 brane-antibrane pair collide and annihilate into ordinary radiation; they do not change $\rho+p_k$ at all since neither branes nor radiation exert pressure in the compact directions and the energy density remains the same.  Yet, after inflation is over and the equation-of-state increases 
to 
$w=+1/3$ (the radiation 
epoch), the NEC violation must be 
reduced or eliminated  to continue
to satisfy the Einstein equations. 
This 
suggests some back-reaction effect must be built into the 
higher dimensional 
theory that creates and later eliminates exponentially 
large NEC-violating 
contributions at the beginning and end of inflation, 
leaving behind exponentially small 
NEC-violating effects needed for the current dark energy dominated 
epoch.   This needs to be incorporated into any realistic 
theory of reheating.\cite{Kofman:2005yz}

The added complexity is disappointing.  Inflation and dark energy in 4d have 
always had the problem
that they require special degrees of freedom and 
fine-tuning.  One would have hoped that extra dimensions, which are 
introduced to simplify the unification of fundamental forces, would 
also alleviate the conditions needed for inflation.  The no-go 
theorems say the opposite: the number and complexity 
of conditions needed to have inflation or dark increase significantly. 

The fact that NEC violation is required to have inflation in theories 
with extra dimensions is unexpected since this was not a requirement 
in the original inflationary models based on four dimensions only.  
Curiously, a criticism raised at times about models with bounces from a contracting phase to an expanding phase, such as the  
ekpyrotic\cite{Khoury:2001wf,Buchbinder:2007ad} and cyclic\cite{Steinhardt:2002ih} alternatives to inflationary cosmology,  is that the bounce requires a violation of the NEC (or quantum gravity
corrections to GR as the FRW scale factor
 $a(t) \rightarrow 0$ that serve the same function). 
  Now we see that, although the details are different, 
all of these cosmologies require NEC violation when incorporated into 
theories with extra dimensions.

In general, 
the no-go theorems are powerful because they span a broad sweep of theories.  
They say that one should be wary
of focusing on one localized region of the extra-dimensions, such as a 
warped throat, since there are non-trivial global constraints.
Second, just because some elements appear to add to the vacuum energy 
or provide an inflaton potential in the 4d effective theory does not 
mean the theory is viable; they may force unacceptable conditions in 
the higher dimensional theory.  Thirdly,  
the NEC violation must be time-varying, 
at least for the class of 
metrics  considered here.
This power of
the no-go theorems derives from the fact that they 
arise from
``macro-to-micro" approach in which the analysis only relies on known
macroscopic properties, although this also means that 
they tell 
us nothing directly about the detailed microphysics needed  
to satisfy or evade them.  

We note that, thus far,
 we have restricted the analysis to no-go 
theorems that are simple to express and simple to prove.  There are 
numerous other relations that must be satisfied to have cosmic 
acceleration that will be considered in future work.
However, we hope the examples shown here and in 
Ref.~\cite{Wesley:2008fg} suffice to show how these no-go 
theorems can be remarkably 
informative, complementing other ways of thinking about how to 
construct higher 
dimensional models.

We would like to thank Daniel Baumann, Alex Dahlen, 
Igor Klebanov, Jean-Luc Lehners and Juan Maldacena
for helpful discussions. 
This work is supported in part by
the US Department of Energy grant DE-FG02-91ER40671.

\newpage 

\appendix

\section{Appendix: Some Useful Relations}

For ${\cal R}${\it -flat} (RF) models, we have the following 
useful relations in 
the case of fixed $\xi$ (breathing mode) and metric $g_{mn}$:
\begin{eqnarray} \label{usefulrf}
G_{00} & =  &-3 \Delta \Omega  -6 (\partial \Omega)^2 + 
e^{\phi-2 \Omega} \, \rho_{4d} \label{firstUsefulRF} \\
p_3 & = & 3 \Delta \Omega - 6 (\partial \Omega)^2 + e^{\phi-
2 \Omega} \, p_{4d} \\
p_k &  =  & (4 - \frac{4}{k}) \Delta \Omega + (10- \frac{4}{k}) 
(\partial \Omega)^2 \\ & &  + e^{\phi-2 \Omega} (
\frac{1}{2} \rho_{4d} (3 w -1 )) \\
\rho + p_3 &  = & e^{\phi-2 \Omega} (\rho_{4d} + p_{4d}) \\
\rho + p_k & = &  (1 - \frac{4}{k}) \Delta \Omega + (4 - 
\frac{4}{k}) (\partial 
\Omega)^2 + e^{\phi-2 \Omega} (\frac{1}{2} \rho_{4d} (1 + 3 w)) \label{appRhoPlusPk}\label{lastUsefulRF}
\end{eqnarray}
manifold.  
Taking $A$-averages and using 
$\langle \Delta \Omega \rangle_A = -A (\partial \Omega)^2$ , we can 
obtain an expression for the equation-of-state of the compact 
directions
\begin{equation} \label{wRF}
w_k^{RF}(A)  = \frac{[(10 - 4A)+ \frac{4A -4}{k}] (\partial \Omega)^2 
+ 
\left( \frac{3 w-1}{2} \right)
e^{\phi}\langle e^{-2 \Omega} \rho_{4d} \rangle_A 
}
{(3 A-6) (\partial \Omega)^2 + e^{\phi}
\langle e^{-2 \Omega} 
\rho_{4d}\rangle_A}.
\end{equation}
Following Ref. \cite{Wesley:2008fg}, we can obtain the RF analogue of (\ref{rangeA}) by multiplying both sides of (\ref{appRhoPlusPk}) by $e^{2\Omega -\phi}$ and taking the $A$-average.  As shown in \cite{Wesley:2008fg}, for all $k \ge 1$ an $A$ can be found such that $A$-dependent coefficients are non-positive.  The RF version of $A_*$, for which the warp term contribution to $e^{-\phi}\langle e^{2\Omega} (\rho+P_k)\rangle_A$ vanishes, is given by
\begin{equation}
A_* = \frac{2(k+2)}{k-4}
\end{equation}
and $A$ can be chosen equal to $A_*$ for all $k \ge 1$.  The choice $A=2$ is inconsistent with keeping the $A$-dependent coefficients non-positive in the RF case.

For {\it conformally} ${\cal R}${\it -flat}                                                                                              
(CRF) models, the analogous relations to (\ref{firstUsefulRF})-(\ref{lastUsefulRF})
are:
\begin{eqnarray} \label{usefulcrf}
G_{00} & =  &(k-4) \Delta \Omega + \frac{1}{2} (k^2 - 3 k - 10) 
(\partial \Omega)^2 + 
e^{\phi-2 \Omega} \, \rho_{4d} \\
p_3 & = & -(k-4) \Delta \Omega - \frac{1}{2} (k^2 - 3 k - 10) 
(\partial \Omega)^2 + e^{\phi-
2 \Omega} \, p_{4d} \\
p_k &  =  & (7 - \frac{6}{k} - k) \Delta \Omega + (6- \frac{2}{k} + 
\frac{5 k}{2} - 
\frac{k^2}{2}) (\partial \Omega)^2 \\ & &  + e^{\phi-2 \Omega} (
\frac{1}{2} \rho_{4d} (3 w -1 )) \\
\rho + p_3 &  = & e^{\phi-2 \Omega} (\rho_{4d} + p_{4d}) \\
\rho + p_k & = &  (3 - \frac{6}{k}) \Delta \Omega + (k+1 - 
\frac{2}{k}) (\partial 
\Omega)^2 + e^{\phi-2 \Omega} (\frac{1}{2} \rho_{4d} (1 + 3 w)) \\
\stackrel{\circ}{R} & =  & 2 (k-1)\, \Delta \Omega + (k-1)(k-2) 
(\partial \Omega)^2,
\end{eqnarray}
where $\stackrel{\circ}{R}$ is the Ricci curvature of the compact 
manifold.  Then, the effective equation-of-state is
\begin{equation} \label{wCRF}
w_k^{CRF}(A)  = \frac{[-(7- \frac{6}{k} -k)A +  
(6 - \frac{2}{k}+ 
\frac{5k}{2} - \frac{k^2}{2})] (\partial \Omega)^2 
+ 
\left(\frac{3 w-1}{2}\right)
e^{\phi}\langle e^{-2 \Omega} \rho_{4d} \rangle_A 
}
{[-(k-4)A + \frac{1}{2}(k^2- 3k -10)] (\partial \Omega)^2 + e^{\phi} \langle 
e^{-2 \Omega} \rho_{4d}\rangle_A}.
\end{equation}

In addition, the following Lemma proven in Ref.~\cite{Wesley:2008fg} 
is useful in 
deriving dark energy theorems:

\noindent
{\it Lemma A1:}  For real and non-zero $\alpha$ and $\beta$, 
\begin{eqnarray}
\alpha \Delta \Omega + \beta (\partial \Omega)^2 & = & \Gamma e^{- 
\gamma \Omega}
\Delta e^{ \gamma \Omega}.
\end{eqnarray}
where $\alpha = \Gamma \gamma$ and $\beta = \Gamma \gamma^2$.  
  

\begin{thebibliography}{test}
\bibitem{note}
In the no-go theorems presented in this paper, the 
term ``compactified models" is short for 
models satisfying the GR, flatness, boundedness and metric conditions 
discussed in Sec.~\ref{assumptions}.
\end{thebibliography}
\end{document}